%% file: main.tex
\documentclass[sigconf, nonacm]{acmart}

\renewcommand\footnotetextcopyrightpermission[1]{}

\acmConference[AgenticSE @ KDD '26]{Agentic Software Engineering (SE 3.0): The Rise of AI Teammates, KDD 2026 Workshop}{August 9, 2026}{Jeju, Republic of Korea}
\acmBooktitle{Agentic Software Engineering (SE 3.0): The Rise of AI Teammates, KDD 2026 Workshop, August 9, 2026, Jeju, Republic of Korea}
\acmYear{2026}
\copyrightyear{2026}

\usepackage{booktabs}
\usepackage{multirow}
\usepackage{graphicx}
\usepackage{subcaption}
\usepackage{amsmath}
\usepackage{enumitem}
\usepackage{xcolor}
\usepackage{listings}
\usepackage{url}

\newcommand{\anonInstitution}{University of Washington}
\newcommand{\anonCenter}{Scientific Software Engineering Center}
\newcommand{\systemName}{\textit{Aleena}}

\title{\systemName{}: Alignment Agent for Research Software Engineering Collaborations}

\author{Kshitij Dani}
\orcid{0009-0006-2706-8808}
\affiliation{
  \institution{eScience Institute}
  \institution{University of Washington}
  \city{Seattle}
  \state{Washington}
  \country{USA}
}
\email{ksdani@uw.edu}

\author{Cordero Core}
\orcid{0000-0002-3531-3221}
\affiliation{
  \institution{eScience Institute}
  \institution{University of Washington}
  \city{Seattle}
  \state{Washington}
  \country{USA}
}
\email{cdcore@uw.edu}

\author{Landung Setiawan}
\orcid{0000-0002-1624-2667}
\affiliation{
  \institution{eScience Institute}
  \institution{University of Washington}
  \city{Seattle}
  \state{Washington}
  \country{USA}
}
\email{landungs@uw.edu}

\author{Carlos Garcia Jurado Suarez}
\orcid{0009-0007-3248-7676}
\affiliation{
  \institution{eScience Institute}
  \institution{University of Washington}
  \city{Seattle}
  \state{Washington}
  \country{USA}
}
\email{carlosg@uw.edu}

\author{Anshul Tambay}
\orcid{0009-0004-9010-1223}
\affiliation{
  \institution{eScience Institute}
  \institution{University of Washington}
  \city{Seattle}
  \state{Washington}
  \country{USA}
}
\email{anshul37@uw.edu}

\author{Vani Mandava}
\orcid{0000-0003-3592-9453}
\affiliation{
  \institution{eScience Institute}
  \institution{University of Washington}
  \city{Seattle}
  \state{Washington}
  \country{USA}
}
\email{vani1@uw.edu}

\author{Anant Mittal}
\orcid{0009-0000-9085-8446}
\affiliation{
  \institution{eScience Institute}
  \institution{University of Washington}
  \city{Seattle}
  \state{Washington}
  \country{USA}
}
\email{anmittal@uw.edu}

\begin{document}

\hyphenpenalty=10000
\tolerance=2000
\emergencystretch=10pt

\raggedbottom

\begin{abstract}

\input{sections/0-abstract}

\end{abstract}




\maketitle

\input{sections/1-introduction}

\input{sections/2-problem}

\input{sections/3-aleena}

\input{sections/4-scenarios}

\input{sections/5-futurework}

\input{sections/6-acknowledgement}


\bibliographystyle{ACM-Reference-Format}
\bibliography{references}

\input{appendix}

\end{document}

%% file: sections/0-abstract.tex
Research software collaborations span meetings, informal chats, pull requests, and GitHub issues. A decision surfaced in a Slack thread, refined in a meeting, and implemented in a pull request can lose its original rationale across these artifacts, leaving domain researchers and research software engineers with divergent mental models of project intent, ownership, and scientific assumptions. We argue that alignment in research software engineering is a continuous lifecycle problem, and that agentic AI can support stakeholder alignment and project-state tracking without replacing human decision-making. We present \systemName{}, an open-source lifecycle alignment agent that uses GitHub as a shared collaboration surface, transforming multi-modal stakeholder interactions into structured project records that surface risks, track open questions, and preserve decision continuity. Grounded in university-based research software engineering center experiences, this paper presents the motivating problem, system design, prototype, and illustrative lifecycle scenarios for \systemName{}.

%% file: sections/1-introduction.tex
\section{Introduction}

Research software collaborations play a central role in advancing science by supporting researchers to translate domain-specific goals, workflows, and computational methods into usable software systems. Yet this software begins as \textit{research prototypes} that prioritize scientific results over engineering practice, and is written by scientists with limited training in software engineering~\cite{johansonSoftwareEngineeringComputational2018, heatonClaimsUseSoftware2015, arvanitouSoftwareEngineeringPractices2021, hocquetnatcom2024, hunter-zinckTenSimpleRules2021, hannayHowScientistsDevelop2009, carverSoftwareEngineeringComputational2012, sandersDealingRiskScientific2008, nieuwpoortDefiningRolesResearch2023}. Compounding this gap, such collaborations operate under evolving scope and incomplete requirements: scientific software developers rarely produce formal requirements documents, because requirements emerge as both the software and the researcher's understanding of the domain mature~\cite{bajraktariRequirementsEngineeringResearch2024}.

In practice, project context is distributed across interaction modalities that are central coordination mechanisms in software engineering teams, including synchronous meetings, asynchronous chats (e.g., Slack or Teams), email threads, and design discussions~\cite{strayUnderstandingCoordinationGlobal2020}. It is further distributed across collaboration platforms such as GitHub, where issues, pull requests, and discussions carry overlapping but partial fragments of project history~\cite{hataGitHubDiscussionsExploratory2022, tsayInfluenceSocialTechnical2014, gousiosExploratoryStudyPullbased2014}. Across these surfaces, two directions of teaching occur in parallel: researchers teach engineers the scientific domain (e.g., data semantics, valid input ranges, modeling assumptions, domain conventions), while engineers teach researchers software practices (e.g., version control, testing, continuous integration, deployment, reproducibility), a translation that depends on \textit{boundary objects} that travel between communities of practice while preserving meaning on each side~\cite{starInstitutionalEcologyTranslations1989, leeBoundaryNegotiatingArtifacts2007}. For mature projects, this distributed context also accumulates legacy decisions, prior design tradeoffs, and rationale behind historical technical choices. 

The consequence is a persistent \textit{alignment} challenge. Researchers describe needs in domain-specific vocabulary, that research software engineers (RSEs) then translate into technical requirements, implementation plans, and maintainable systems. The challenge is not only capturing \textit{what} the team decided, but preserving \textit{why} decisions changed as the collaboration progressed, a long-standing problem of design rationale capture in software engineering~\cite{burgeSoftwareEngineeringUsing2008, conklinGIBISHypertextTool1988, hornerEffectiveDesignRationale2006}. When that reasoning becomes difficult to reconstruct, collaborators lose the \textit{shared mental model} on which coordinated team work depends~\cite{salasSharedMentalModels1993, espinosaFamiliarityComplexityTeam2007}.

This challenge is particularly acute for research software engineering centers such as the \anonCenter{} (SSEC), whose engagements are typically short and interdisciplinary. Through a retrospective activity following the delivery of more than 20 projects with multiple stakeholders across multiple organizations, our center identified recurring alignment challenges spanning \textit{stakeholder expectations}, \textit{scientific vocabulary and domain assumptions}, \textit{shifting ownership and review responsibilities}, and \textit{decision continuity} (Section~\ref{sec:problem})~\cite{tambaySSECYearOneLearings2024}. Together, these challenges suggest that alignment is a continuous problem of communication, collaboration, and coordination across the research software collaboration lifecycle.


To address this problem, we designed and developed \systemName{}, an open-source \textit{lifecycle alignment agent}. \systemName{} treats GitHub as a central collaboration surface and operates across multi-modal project artifacts, including meeting and chat transcripts, GitHub issues, discussions, and pull requests. The current prototype transforms submitted artifacts into structured \textit{summaries}, \textit{issues}, \textit{risks}, \textit{open questions}, and \textit{suggestions} that help stakeholders act on emerging alignment concerns. More broadly, \systemName{} is designed to \textit{support} alignment without replacing human decision-making, helping team members preserve context and maintain shared project understanding. In framing \systemName{} as a project-state-aware agent that collaborates with rather than supplants human stakeholders, we contribute to the emerging vision of SE~3.0 in which AI teammates participate in software work alongside human developers~\cite{liRiseAITeammates2025}.

This paper makes two contributions: \textbf{(i)}~we reframe alignment in research software engineering as a continuous project-state management problem rather than a one-time scoping activity, connecting stakeholder expectations, scientific terminology, ownership responsibilities, and decision continuity across the lifecycle; and \textbf{(ii)}~we present \systemName{}, an open-source lifecycle alignment agent, describe its prototype, and illustrate how it addresses alignment concerns across project phases, from artifact ingestion (e.g., meetings, chats, issues, pull requests) to structured GitHub records (e.g., summaries, risks, open questions, terminology-drift signals, draft pull requests).

%% file: sections/2-problem.tex
\section{Problem Setting}
\label{sec:problem}
The ~\anonCenter{} (SSEC) at the ~\anonInstitution{} provides research software engineering support to scientists who build, improve, or sustain software for research. Collaborators come from diverse domains such as seismology, genetics, and oceanography, bringing scientific goals, domain expertise, and established workflows, but not always the engineering capacity to turn those goals into maintainable, deployable tools. Engagements typically last three to six months, requiring researchers and engineers to establish shared understanding quickly. The center's role is therefore not only to write code, but to translate scientific needs into software that remains useful and maintainable beyond the immediate collaboration.

\begin{figure}[t]
    \centering
    \includegraphics[width=\linewidth]{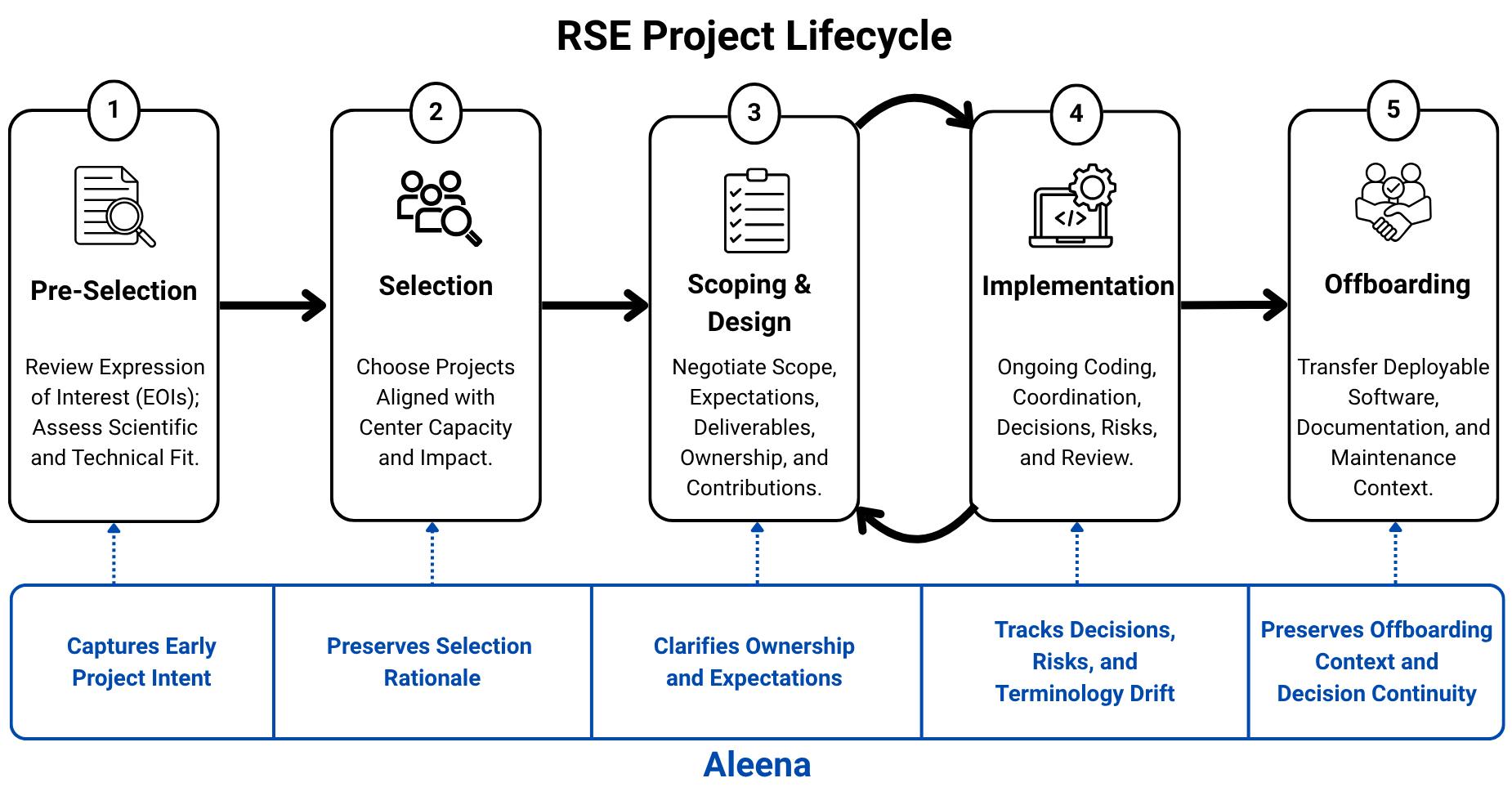}
    \caption{The RSE project lifecycle at and \systemName{}'s role across it. The top row shows the five phases (Pre-Selection, Selection, Scoping and Design, Implementation, Offboarding); the bottom row summarizes how \systemName{} supports each phase. Together they ground the four alignment challenges AC1--AC4 that recur throughout the lifecycle.}
    \Description{Horizontal infographic. Top row: five numbered, arrow-connected boxes for the RSE project phases (Pre-selection, Selection, Scoping and Design, Implementation, and Offboarding), each with an icon and a short description of its activities. Bottom row: a banner labeled ``Aleena'' with five items aligned beneath the phases, describing \systemName{}'s supporting role at each step (capturing early project context, preserving selection rationale, clarifying expectations and ownership, tracking decisions and terminology drift, and preserving offboarding context and decision continuity).}
    \label{fig:rse_project_lifecycle}
\end{figure}

As shown in Figure~\ref{fig:rse_project_lifecycle}, RSE projects move through five phases. \textit{Pre-Selection} and \textit{Selection} review proposals from principal investigators (PIs) and draw on scientific expertise to assess feasibility, translating an initial framing into a scopable collaboration. In \textit{Scoping and Design}, researchers and RSEs negotiate technical boundaries (requirements, architecture, MVP deliverables, out-of-scope features) and sociotechnical commitments (RSE, PI, and research team responsibilities), surfacing alignment challenges around timelines, time commitments, ownership, and proposed engineering solutions. During the \textit{Implementation}, alignment must be actively maintained as assumptions and decisions evolve~\cite{storeyTechnicalDebtCognitive2026}. Weekly meetings, GitHub issues, pull requests, chats, and design documents become the primary artifacts through which direction and scope is refined, but the reasoning behind decisions becomes distributed across them, making it difficult to connect new requirements to original goals or to ensure a shared understanding of project direction.

A related challenge is preserving scientific fidelity while making engineering decisions. Scientific domains carry specialized vocabularies and assumptions that are often unfamiliar to RSEs~\cite{gesingRSEs2035Surviving2025, mcinnesReport2025Workshop2025}, making it difficult to interpret researcher requirements or to recognize how a technical change may alter scientific meaning. Researchers, in turn, need visibility into the engineering process to judge whether the implementation still supports the science. Bridging these perspectives requires \textit{boundary-negotiating artifacts} that travel between scientific and engineering communities of practice while preserving meaning on each side~\cite{starInstitutionalEcologyTranslations1989, leeBoundaryNegotiatingArtifacts2007}. Ownership and review therefore shift across the lifecycle: tasks affecting scientific meaning (transforming data, defining valid input or output ranges, implementing analysis logic) may require scientific review, while engineering tasks (packaging, continuous integration, version control, infrastructure) are typically led by engineers.

By the \textit{Offboarding} phase, the goal is to deliver software that researchers can continue to use and maintain after the engagement ends. This requires more than a final handover of code and documentation: throughout implementation, researchers must also build familiarity with the codebase, contribution practices, and maintenance expectations needed to sustain the project~\cite{phamOnboardingInexperiencedDevelopers2017}. Offboarding therefore depends on whether technical outputs, supporting materials, and collaboration practices are sufficiently structured for continued use, maintenance, and community contribution.

Across this lifecycle, four alignment challenges (AC) are especially important:
\begin{description}[leftmargin=*, itemsep=2pt, topsep=2pt]
    \item[\textbf{AC1.\ Stakeholder Expectations.}] Expectations must remain explicit, including what the center will build and what researchers are expected to contribute.
    \item[\textbf{AC2.\ Scientific Vocabulary and Domain Assumptions.}] Researchers and RSEs must stay aligned on domain terminology and assumptions in order to act on shared meaning.
    \item[\textbf{AC3.\ Ownership and Review Responsibilities.}] Ownership must remain visible as tasks shift between work led by researchers, led by engineers, or jointly owned.
    \item[\textbf{AC4.\ Decision Continuity.}] Evolving assumptions, decisions, and expectations are often scattered across project artifacts. Preserving this context helps future maintainers understand why decisions were made and supports maintenance, offboarding, and community contribution.
\end{description}

These challenges motivate \systemName{}, a lifecycle alignment agent that keeps stakeholder expectations explicit, aligns scientific and engineering terminology, makes ownership and review visible, and preserves evolving assumptions and decisions across project artifacts, all without removing agency from human stakeholders.

%% file: sections/3-aleena.tex
\section{\systemName{}}
\label{sec:aleena}

We created \textbf{\systemName{}}, an open-source lifecycle alignment agent built on the premise that alignment is not a single outcome of scoping but a project state that must be continuously maintained. \systemName{} maintains an evolving project-state model organized around AC1--AC4, ingesting meeting and chat transcripts, GitHub issues, discussions, and pull requests alongside historical artifacts (prior design documents, archived discussions, issue histories) that preserve the rationale behind earlier decisions. Grounded in this accumulated context, it identifies friction points and generates recommendations tailored to the collaboration while leaving decision authority with human stakeholders~\cite{bridgefordTenSimpleRules2025, obrienThreatsScientificSoftware2025, wuHumangenerativeAICollaboration2025}.

\paragraph{Agentic structure.}
The \systemName{} agent operates across an explicit four-step loop. \textbf{(1)~Perceive.} It ingests stakeholder-submitted artifacts together with prior project history retrieved from the repository. \textbf{(2)~Update.} It extracts structured signals (action items, risks, open questions, terminology-drift candidates, ownership transitions) and merges them into the project-state model. \textbf{(3)~Select.} It chooses an action from a constrained set of GitHub-native operations (open issue, open discussion, draft pull request, post comment, cross-link to a prior artifact), grounded in the current project state. \textbf{(4)~Defer.} It stops short of any merge, close, or external action; human stakeholders accept, edit, or reject every output. This structure preserves human agency and externalizes decision rationale, positioning \systemName{} within the broader class of LLM-based autonomous agents that select tools and actions over external environments~\cite{wangSurveyLargeLanguage2024, schickToolformerLanguageModels2023, yangSWEagentAgentComputerInterfaces2024, jimenezSWEbenchCanLanguage2023} while staying anchored to the SE~3.0 vision of AI teammates collaborating with human developers~\cite{liRiseAITeammates2025}. By representing alignment concerns where implementation and technical discussion already occur~\cite{tsayInfluenceSocialTechnical2014, gousiosExploratoryStudyPullbased2014}, \systemName{} turns unclear expectations, emerging risks, ambiguous terminology, unclear ownership, and fragmented decisions into actionable GitHub artifacts, continuing a line of work on bots and meeting summarization that mediates developer collaboration through structured artifacts~\cite{storeyDisruptingDeveloperProductivity2016, goliaActionItemDrivenSummarizationLong2024, asthanaSummariesHighlightsAction2025}.

\paragraph{End-to-end workflow.}
\begin{figure}[tb]
    \centering
    \includegraphics[width=\linewidth]{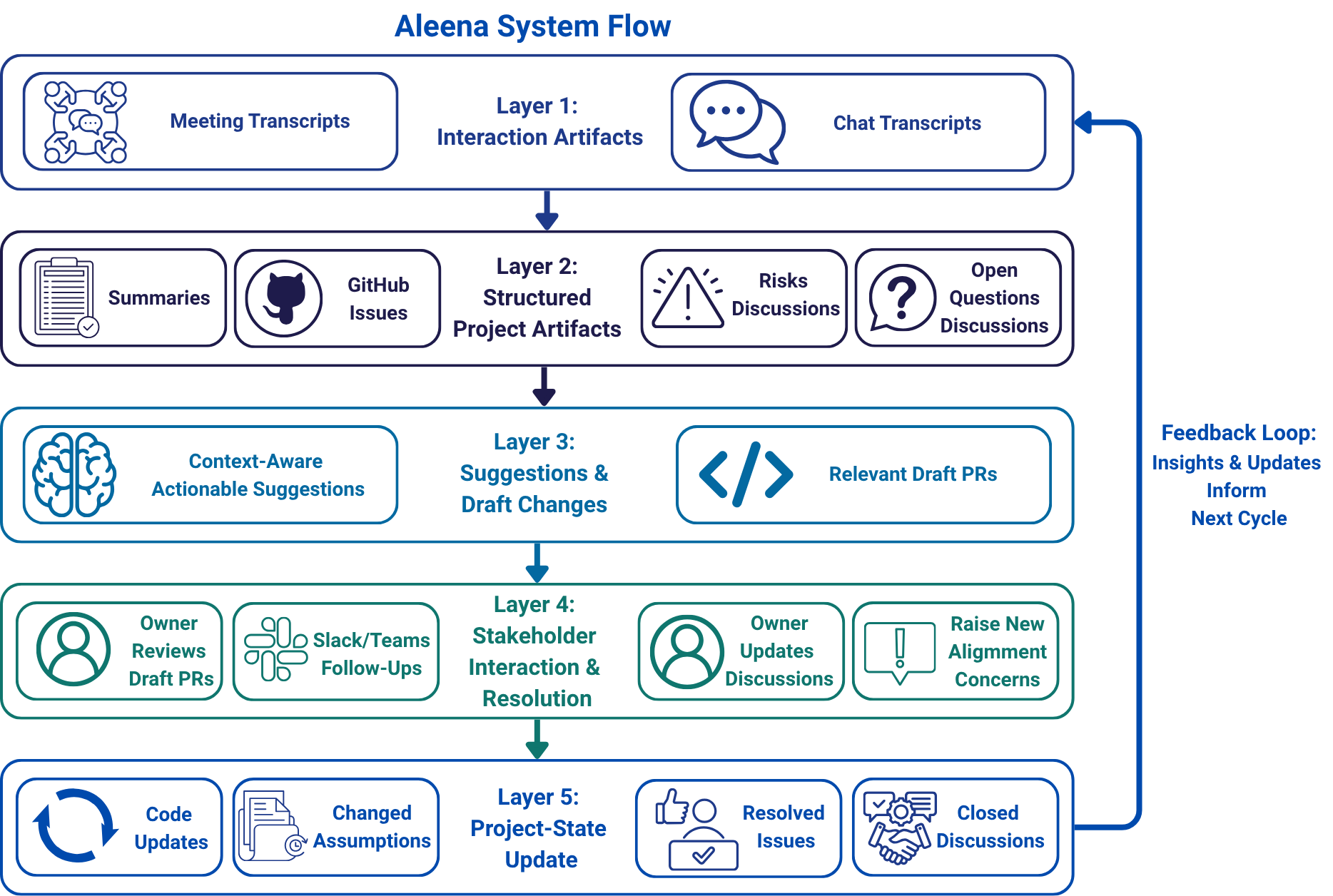}
    \caption{End-to-end \systemName{} workflow, organized as five layers that move from interaction artifacts to project-state updates and back. Layer~1 captures meeting and chat transcripts; Layer~2 transforms them into structured GitHub artifacts (meeting summaries, issues, risks, and discussions); Layer~3 produces context-aware suggestions and draft pull requests grounded in prior interaction context; Layer~4 routes these outputs through stakeholder review and follow-up; and Layer~5 closes the loop by updating the project state ahead of the next meeting, feeding insights back into Layer~1.}
    \Description{Vertical, five-layer flow diagram titled ``\systemName{} System Flow.'' Layer 1, ``Interaction Artifacts,'' contains Meeting Transcripts and Chat Transcripts. Layer 2, ``Structured Project Artifacts,'' contains Meeting Summary, GitHub Issues, Risk Discussions, and Open Question Discussions. Layer 3, ``Suggestions and Draft Changes,'' contains Actionable Suggestions, Relevant Draft PRs, Context-Aware Suggestions, and Relevant Prior Interaction Context. Layer 4, ``Stakeholder Interaction and Resolution,'' contains Owner reviews draft PR, Risk and open question discussions continue, Slack/Teams follow-ups can be re-submitted to \systemName{}, \systemName{} updates discussions, and Raise new alignment concerns or close resolved items. Layer 5, ``Next Meeting and Project State Update,'' contains Code updates, Changed assumptions, Resolved issues/open questions, Close completed issues and discussions, and Raise new alignment concerns or risks. A feedback loop on the right connects Layer 5 back to Layer 1, labeled ``Feedback loop: Insights and updated context inform the next cycle.''}
    \label{fig:aleena_system_flow}
\end{figure}

Figure~\ref{fig:aleena_system_flow} summarizes the prototype as a five-layer loop; Layer~5's feedback into Layer~1 distinguishes \systemName{} from a one-shot summarizer and lets it treat alignment as continuously maintained. The loop is intentionally user-driven: artifacts enter the project context only when a stakeholder submits them, in line with ethical and open-science guidance on generative AI in research workflows~\cite{hofeditzEthicalChallengesHuman2024, hosseiniOpenScienceGenerative2024}.

\paragraph{Prototype design and implementation.}
\begin{figure}[tb]
    \centering
    \includegraphics[width=\linewidth]{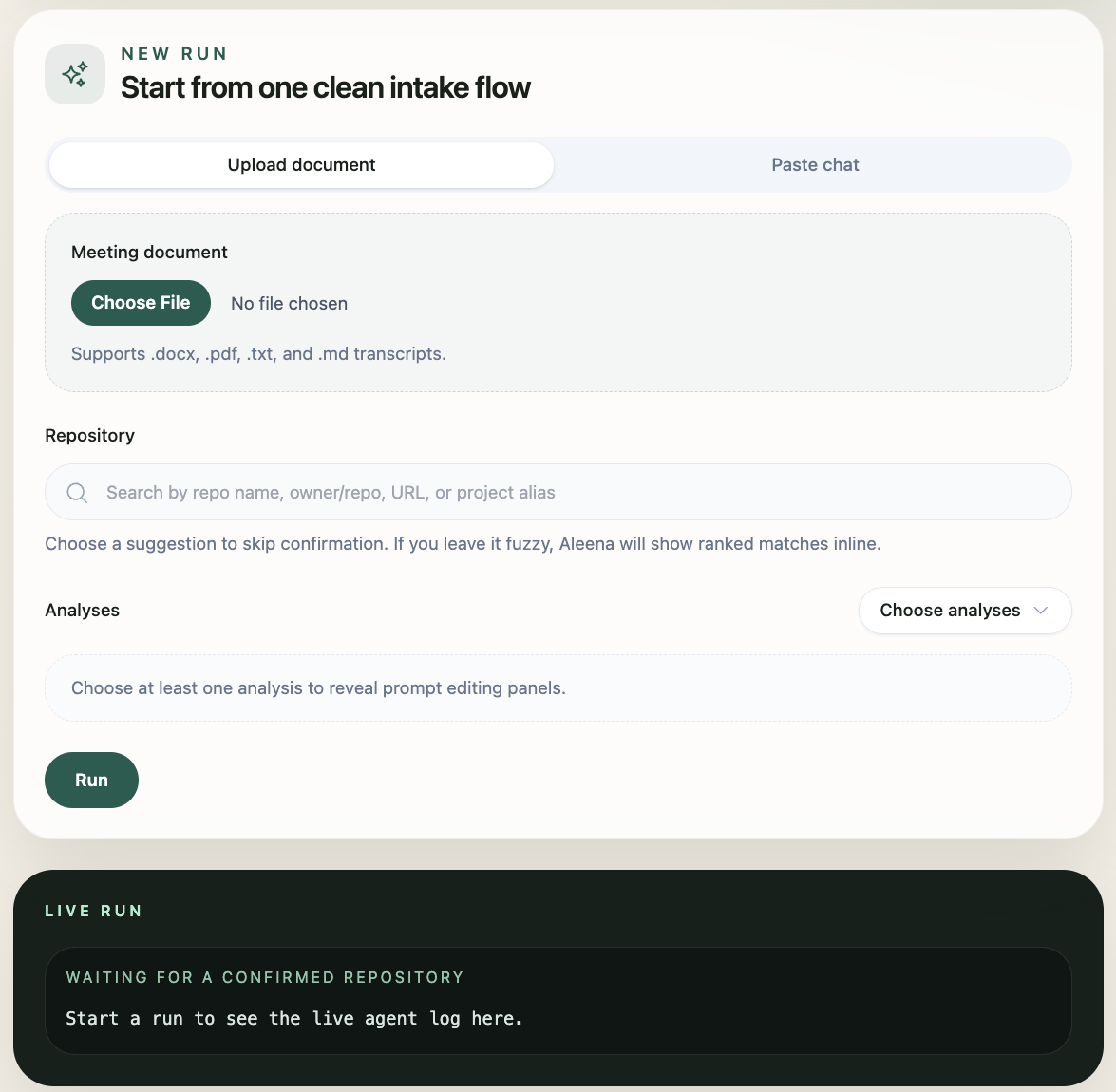}
    \caption{The current \systemName{} prototype interface. Stakeholders sign in with GitHub, upload a meeting document or paste a chat transcript, confirm the target repository, and select which analyses to run. Each analysis exposes an editable prompt so that researchers and RSEs can inspect and adjust how \systemName{} interprets the artifact before it produces GitHub-native outputs. A live run log makes the agent's progress observable rather than opaque.}
    \Description{Screenshot of the \systemName{} web application. A header shows the \systemName{} title, a one-line description (``Upload transcripts, match them to the right repository, and turn meetings into concrete engineering follow-through.''), the signed-in user, and a sign-out control. Below, a ``New Run'' card offers two tabs (``Upload document'' and ``Paste chat'') and a file picker that accepts .docx, .pdf, .txt, and .md transcripts. A repository search field invites the user to find a target repository by name, URL, or project alias. An analyses dropdown is set to ``Vocabulary Analysis + Meeting minutes, action items, risks, and open questions.'' Two collapsible panels expose the Vocabulary Analysis prompt and the Meeting extraction prompt for editing. A green ``Run'' button starts the analysis. A dark ``Live Run'' panel at the bottom shows a placeholder log waiting for a confirmed repository.}
    \label{fig:aleena_prototype_v2}
\end{figure}

The \systemName{} prototype, shown in Figure~\ref{fig:aleena_prototype_v2}, is a web application (FastAPI backend, ReactJS frontend) that coordinates artifact submission, AI gateway calls, and GitHub API writes. A GitHub OAuth App handles user sign-in and a GitHub App provides repository-scoped installation, so \systemName{} operates only within the permissions granted on a center's project repositories rather than across the organization. Submitted artifacts flow through the center's LiteLLM-based AI gateway, which routes requests across multiple LLM providers behind a common interface based on availability, privacy, and cost~\cite{liLLMRouterBenchMassiveBenchmark2026, fengLLMRouterOpenSourceLibrary2025}. The backend is organized around analyzer modules, each responsible for a specific alignment task (e.g., extracting meeting summaries, detecting vocabulary drift). Each analyzer sends the artifact to the AI gateway with a task-specific prompt and structured output examples that, as Figure~\ref{fig:aleena_prototype_v2} shows, stakeholders can review or refine before a run. New analyzers can be added as additional alignment challenges emerge, consistent with recent arguments for specialized rather than monolithic models in agentic systems~\cite{belcakSmallLanguageModels2025}, and their outputs are written through the GitHub API as discussions, issues, comments, and, where appropriate, draft pull requests.

\paragraph{Positioning relative to other platforms.}
Within the broader agentic ecosystem, systems such as OpenClaw~\cite{OpenClawPersonalAI}, Hermes~\cite{NousResearchHermesagent2026}, and Paperclip~\cite{PaperclipAppPeople} target personal productivity, adaptive personal assistance, and multi-agent orchestration respectively, and are oriented toward taking action (sending mail, executing code, deploying pull requests, managing budgets) on behalf of an individual principal. Platforms such as Jira ~\cite{Jira} and Notion ~\cite{Notion} help teams organize action items, documentation and project state, but do not reason whether the team remains aligned across goals, constraints, responsibilities and decisions.

\systemName{} occupies a distinct layer: its principal is a multi-stakeholder research collaboration, its state is the alignment surface of a project (AC1--AC4) rather than user memory or an agent organization chart, and it is constrained to \emph{not} decide on behalf of researchers or RSEs. It is therefore complementary: a Paperclip-orchestrated fleet, an OpenClaw, or a Hermes deployment could each submit artifacts \emph{into} \systemName{}, whose outputs remain GitHub records intended for human review.

\paragraph{Deployment status and scope.}
\systemName{} is deployed at the \anonCenter{} and has begun seeing early operational use on a number of active collaborative software projects, where stakeholders have submitted meeting and chat transcripts and reviewed the resulting GitHub issues, discussions, and draft pull requests. This paper reports on \systemName{}'s design and illustrative lifecycle scenarios (Section~\ref{sec:scenarios}) as we build history to evaluate and measure, per engagement and per lifecycle phase, how many transcripts are ingested, how many GitHub artifacts are produced, and what fraction of those artifacts stakeholders \emph{retain} as-is, \emph{edit} before acting on, or \emph{close without action}. 

\paragraph{Future evaluation.}
We are instrumenting \systemName{} to collect metrics that can help substantiate its role as an alignment agent across AC1--AC4. These include the number of GitHub issues, draft pull requests, and discussions generated per project, as well as the fraction of artifacts that stakeholders retain as-is, edit before acting on, or close without action. We will also track cases where closed issues are re-opened, which may indicate unresolved or recurring alignment concerns, and cases where stakeholders correct issue ownership, which provides feedback on whether \systemName{} assigns review responsibility appropriately. For example, when an incorrectly assigned issue is reassigned to the appropriate stakeholder, that correction can be stored as project context to improve future ownership recommendations. These quantitative signals will be complemented by stakeholder interviews to assess whether \systemName{} improves coordination, preserves decision continuity, and surfaces useful alignment concerns during research software collaborations.

%% file: sections/4-scenarios.tex
\section{Alignment Challenges Addressed by \systemName{}}
\label{sec:scenarios}

The three scenarios below trace AC1--AC4 across the RSE project lifecycle (Figure~\ref{fig:rse_project_lifecycle}), from Scoping and Project Design through the Implementation and into Offboarding. Each follows the same path through \systemName{}'s loop: a stakeholder submits an artifact, \systemName{} produces structured GitHub records grounded in prior project context, and human collaborators retain authority over interpretation and action.

\paragraph{Early-Stage Stakeholder Expectations (AC1).}
In the Scoping and Project Design phase, stakeholders hold an initial meeting to define collaboration goals over three to six months: what the center will deliver, what domain input the PI will provide, how responsibilities are distributed, and how success will be evaluated. Even at this early stage, assumptions are often left implicit or interpreted differently across participants. When a stakeholder submits the meeting transcript through \systemName{}, the agent converts it into a structured GitHub artifact capturing initial intent, stakeholder expectations, key assumptions, and areas of responsibility. This artifact (Appendix~\ref{app:meeting-minutes-output}) anchors later decisions to a shared baseline that all stakeholders can access, revisit, and revise as the project evolves, rather than to any one participant's divergent recollection.

\paragraph{Mid-Project Feature Concerns (AC2, AC3).}
During the Implementation phase, stakeholders review an interface developed by the engineering team. A researcher observes that parts of the interface are not intuitive in their scientific domain; for instance, several user-interface labels do not reflect domain vocabulary, and a term like \textit{model} is used inconsistently, referring to a domain-specific simulation in the researcher's framing and to a serialized object in the engineering codebase. \systemName{}'s vocabulary analyzer (Appendix~\ref{app:vocabulary-drift-prompt})  surfaces such conflicting interpretations as a ranked list of candidate terms with their occurrences across artifacts, and opens a GitHub discussion thread that links each occurrence so the team can converge on a shared definition (AC2). For ownership (AC3), \systemName{} tags each issue with the stakeholders whose review the change requires (a researcher when the change affects scientific meaning, e.g., renaming a domain label or changing valid input ranges; an RSE when it affects packaging or infrastructure; and both when the change spans them), drawing on signals from the meeting transcript and prior issue history. For changes that are small and well-scoped, \systemName{} additionally drafts a pull request (Appendix~\ref{app:draft_PR}) that begins addressing the concern; the owner reviews the draft and decides whether to merge, ensuring that authority over scientific meaning and engineering implementation remains with human collaborators.

\paragraph{Decision Continuity Across Interactions (AC4).}
Later in the Implementation phase, and continuing into Offboarding, a project meeting surfaces an unresolved risk about how a new requirement affects the current implementation plan. \systemName{} converts the concern into a GitHub discussion, tags the relevant stakeholders, and offers practical follow-up guidance. As additional artifacts arrive (subsequent meetings, chat threads, related issues), \systemName{} updates the existing GitHub discussion rather than treating each interaction as a separate record. The resulting trail shows how assumptions changed, why the concern was resolved, and how the project moved from uncertainty toward a clearer implementation direction. This continually summarized record also supports efficient retrieval when accumulated project state is reused in later interactions~\cite{ganRAGMCPMitigatingPrompt2025}.

%% file: sections/5-futurework.tex
\section{Discussion and Future Work}
\label{sec:futurework}

Two concerns shape \systemName{}'s intended use. First, privacy, governance, and retention: \systemName{} only ingests meeting and chat artifacts that stakeholders intentionally upload, and does not automatically monitor communication channels. Acceptable transcripts must be approved by participating stakeholders for project use, and access to resulting records is governed by repository-scoped GitHub permissions and the project’s retention practices. When a repository is transferred to a researcher or external owner after handover, \systemName{} may lose access to project context. This motivates an offboarding memory that preserves key decisions, assumptions, and risks before access changes. Second, over-reliance: prior work documents both productivity gains and degradations of skill, vigilance, and trust calibration when humans rely heavily on generative AI for code and decisions, including in scientific computing settings~\cite{obrienThreatsScientificSoftware2025, shenHowAIImpacts2026, wuHumangenerativeAICollaboration2025, kabirStackOverflowObsolete2024, obrienSurveyGenerativeAI2026, obrienHowScientistsUse2025}, so \systemName{} remains a support layer in which final interpretation and action stay with human collaborators. When \systemName{} produces incorrect terminology links, misleading risk summaries, unnecessary issues, or wrong ownership assignments, these artifacts remain reviewable GitHub records that stakeholders can revise, reassign, close, merge, or reject before they become accepted project decisions.

\systemName{} is also designed to be open source and portable, with a GitHub-centered architecture and infrastructure-as-code deployment that lets other organizations host their own instances.

A key next step is to connect interaction artifacts across the lifecycle so \systemName{} can reason about project-wide alignment rather than only interaction-level concerns, detecting broader drift in expectations, ownership, and priorities that compounds otherwise as cognitive and intent debt~\cite{storeyTechnicalDebtCognitive2026, bajraktariRequirementsEngineeringResearch2024}. We aspire to evolve \systemName{} into a conversational lifecycle agent that retrieves prior history and grounds suggestions in collaboration and code context, while remaining mindful that conversational agents lose track of multi-turn context~\cite{labanLLMsGetLost2025} and benefit from retrieval-grounded tool selection~\cite{ganRAGMCPMitigatingPrompt2025}. A planned multi-engagement evaluation will report retention, edit, and close-without-action rates for \systemName{}'s GitHub outputs across the lifecycle phases of Figure~\ref{fig:rse_project_lifecycle}. These measurements complement AIDev~\cite{liRiseAITeammates2025}: where AIDev records what AI teammates produce at scale, \systemName{} studies the upstream alignment context that shapes what they should produce in research software collaborations, contributing an RSE perspective to the broader SE~3.0 agenda.

%% file: sections/6-acknowledgement.tex
\section{Acknowledgements}
\label{sec:6-acknowledgement.tex}

The authors thank the National AI Research Resource (NAIRR) Pilot program, funded by the National Science Foundation, which provided Microsoft Azure and Hugging Face credits (Award \#240292) used to build the SSEC's AI gateway stack. We thank Amazon and AWS Bedrock for the credits provided to the eScience Institute, which enabled \systemName{} to use generative AI models. The University of Washington Scientific Software Engineering Center is supported by Schmidt Sciences as part of their Virtual Institute for Scientific Software program.

%% file: appendix.tex
\appendix

\clearpage
\section{Prompts}
\subsection{Meeting Minutes, Action Items, Risks, and Open Questions}
\label{app:meeting-summary-prompt}

The following prompt is used by \systemName{} to summarize UW SSEC meeting transcripts into a structured JSON format.

\begin{lstlisting}[basicstyle=\ttfamily\small,breaklines=true,frame=single]
You summarize UW SSEC meetings.

Return strict JSON with keys:
- summary: string
- decisions: array of strings
- action_items: array of strings
- risks: array of strings
- open_questions: array of strings

Requirements:
- The summary must be 2-5 sentences and should never be empty.
- Extract at least the most important concrete points present in the text.
- If a section has no content, return an empty array for that section.
- Do not invent facts that are not present in the meeting text.
- Prefer concrete project names, deliverables, blockers, and next steps when available.
- Never return an empty object.

Example JSON:
{
  "summary": "The team reviewed the current project status, clarified blockers, and aligned on next steps.",
  "decisions": ["Use the current transcript ingestion path for testing."],
  "action_items": ["Confirm the target repository before running terminology analysis."],
  "risks": ["The LiteLLM endpoint may still fail if DNS is misconfigured."],
  "open_questions": ["Should meeting summaries always be posted as Discussions?"]
}
\end{lstlisting}

\noindent\begin{minipage}{\linewidth}
\subsection{Vocabulary Extraction}
\label{app:vocabulary-extraction-prompt}

The following prompt is used to extract project-specific terminology from UW SSEC meeting transcripts.

\begin{lstlisting}[basicstyle=\ttfamily\small,breaklines=true,frame=single]
You extract domain terminology from meeting transcripts.
Return strict JSON with this schema:
{
  "terms": [
    {
      "term": "string",
      "definition": "string",
      "evidence": "short quote or paraphrase",
      "confidence": 0.0
    }
  ]
}
Rules:
- Include only meaningful domain/project terms.
- Skip generic words.
- Keep definitions concise.
- If uncertain, omit the term.
- Output raw JSON only.
- Do not use markdown code fences.
- Do not add commentary before or after JSON.
\end{lstlisting}
\end{minipage}
\subsection{Vocabulary Drift}
\label{app:vocabulary-drift-prompt}

The following prompt is used to compare definitions of the same term and identify whether they are compatible or conflicting.

\begin{lstlisting}[basicstyle=\ttfamily\small,breaklines=true,frame=single]
You compare two definitions of the same term.
Return strict JSON only with schema:
{"relation": "duplicate" | "conflict", "reason": "string"}
Use 'duplicate' if definitions are equivalent/compatible, else 'conflict'.
\end{lstlisting}

\clearpage
\section{Images of Outputs}
\label{app:output-images}

\subsection{Meeting Minutes, Action Items, and Risks}
\label{app:meeting-minutes-output}

\begin{figure}[H]
    \centering
    \includegraphics[width=\linewidth]{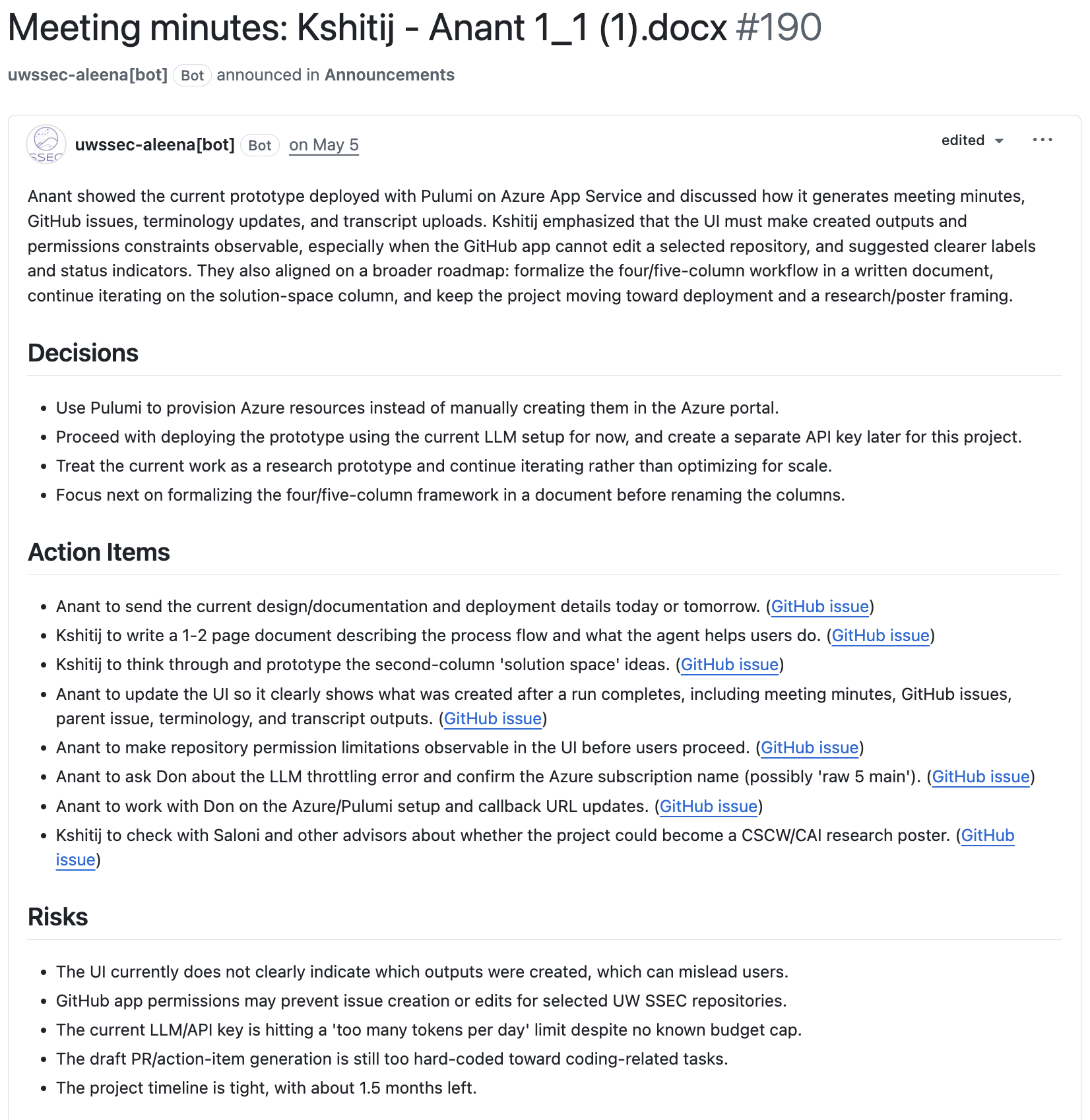}
    \caption{\systemName{} first lists decision taking during the engagement. Next it creates action items as GitHub issues and assigns them to relevant stakeholders. Finally, it generates separate GitHub discussions for risks and links them to the meeting summary discussion as parent or related items.}
    \Description{Screenshot of an Aleena-generated GitHub Discussion showing meeting decisions, action items, and related risks.}
    \label{fig:aleena_meeting_minutes_example}
\end{figure}

Figure~\ref{fig:aleena_meeting_minutes_example} shows an example of how \systemName{} creates a GitHub Discussion that records decisions made during a specific engagement using the prompt in Appendix~\ref{app:meeting-summary-prompt}.

\subsection{Draft PR}
\label{app:draft_PR}

\noindent\begin{minipage}{\linewidth}
Figure~\ref{fig:aleena_draft_PR} shows an example of a draft PR generated by \systemName{} based on an interaction. In this scenario, the owner of the PR reviews, approves, and manually merges the PR.

\medskip

\centering
\includegraphics[width=\linewidth]{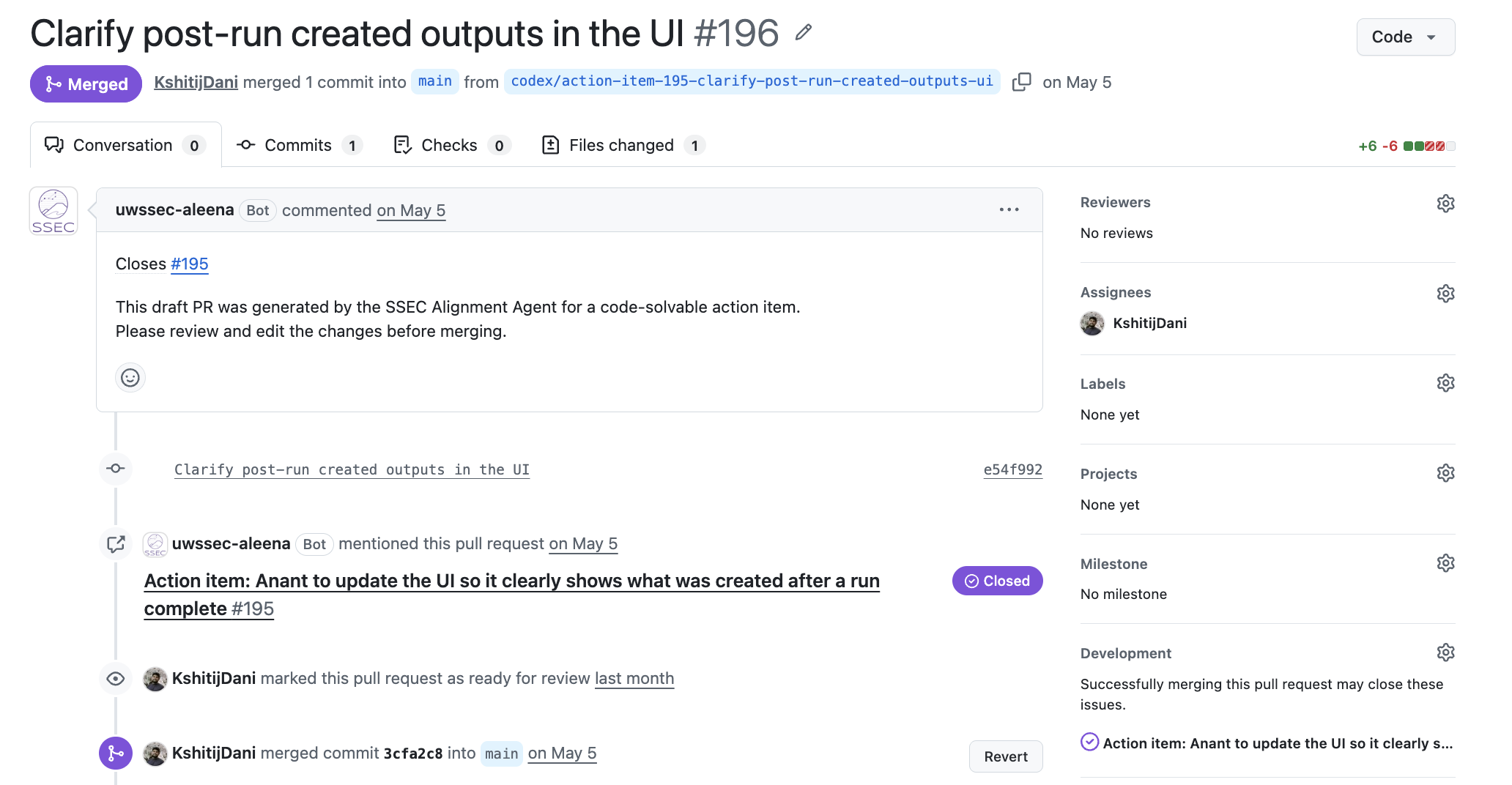}
\captionof{figure}{\systemName{} first determines action items that can be resolved using a draft PR. It then generates a draft PR and assigns it to a relevant owner for manual review and approval.}
\label{fig:aleena_draft_PR}
\end{minipage}

%% file: main.bbl

\begin{thebibliography}{53}


\ifx \showCODEN    \undefined \def \showCODEN     #1{\unskip}     \fi
\ifx \showISBNx    \undefined \def \showISBNx     #1{\unskip}     \fi
\ifx \showISBNxiii \undefined \def \showISBNxiii  #1{\unskip}     \fi
\ifx \showISSN     \undefined \def \showISSN      #1{\unskip}     \fi
\ifx \showLCCN     \undefined \def \showLCCN      #1{\unskip}     \fi
\ifx \shownote     \undefined \def \shownote      #1{#1}          \fi
\ifx \showarticletitle \undefined \def \showarticletitle #1{#1}   \fi
\ifx \showURL      \undefined \def \showURL       {\relax}        \fi
\providecommand\bibfield[2]{#2}
\providecommand\bibinfo[2]{#2}
\providecommand\natexlab[1]{#1}
\providecommand\showeprint[2][]{arXiv:#2}

\bibitem[Arvanitou et~al\mbox{.}(2021)]%
        {arvanitouSoftwareEngineeringPractices2021}
\bibfield{author}{\bibinfo{person}{Elvira-Maria Arvanitou}, \bibinfo{person}{Apostolos Ampatzoglou}, \bibinfo{person}{Alexander Chatzigeorgiou}, {and} \bibinfo{person}{Jeffrey~C. Carver}.} \bibinfo{year}{2021}\natexlab{}.
\newblock \showarticletitle{Software {Engineering} {Practices} for {Scientific} {Software} {Development}: {A} {Systematic} {Mapping} {Study}.}
\newblock \bibinfo{journal}{\emph{Journal of Systems and Software}}  \bibinfo{volume}{172} (\bibinfo{date}{Feb.} \bibinfo{year}{2021}), \bibinfo{pages}{110848}.
\newblock
\showISSN{01641212}
\href{https://doi.org/10.1016/j.jss.2020.110848}{doi:\nolinkurl{10.1016/j.jss.2020.110848}}


\bibitem[Asthana et~al\mbox{.}(2025)]%
        {asthanaSummariesHighlightsAction2025}
\bibfield{author}{\bibinfo{person}{Sumit Asthana}, \bibinfo{person}{Sagi Hilleli}, \bibinfo{person}{Pengcheng He}, {and} \bibinfo{person}{Aaron Halfaker}.} \bibinfo{year}{2025}\natexlab{}.
\newblock \showarticletitle{Summaries, {Highlights}, and {Action} {Items}: {Design}, {Implementation} and {Evaluation} of an {LLM}-{Powered} {Meeting} {Recap} {System}}.
\newblock \bibinfo{journal}{\emph{Proceedings of the ACM on Human-Computer Interaction}} \bibinfo{volume}{9}, \bibinfo{number}{2} (\bibinfo{date}{May} \bibinfo{year}{2025}), \bibinfo{pages}{CSCW176:1--CSCW176:29}.
\newblock
\href{https://doi.org/10.1145/3711074}{doi:\nolinkurl{10.1145/3711074}}


\bibitem[{Atlassian}(2026)]%
        {Jira}
\bibfield{author}{\bibinfo{person}{{Atlassian}}.} \bibinfo{year}{2026}\natexlab{}.
\newblock \bibinfo{title}{Jira}.
\newblock
\urldef\tempurl%
\url{https://www.atlassian.com/software/jira}
\showURL{%
\tempurl}


\bibitem[Bajraktari et~al\mbox{.}(2024)]%
        {bajraktariRequirementsEngineeringResearch2024}
\bibfield{author}{\bibinfo{person}{Adrian Bajraktari}, \bibinfo{person}{Michelle Binder}, {and} \bibinfo{person}{Andreas Vogelsang}.} \bibinfo{year}{2024}\natexlab{}.
\newblock \showarticletitle{Requirements {Engineering} for {Research} {Software}: {A} {Vision}}. In \bibinfo{booktitle}{\emph{2024 {IEEE} 32nd {International} {Requirements} {Engineering} {Conference} ({RE})}}. \bibinfo{publisher}{IEEE}, \bibinfo{address}{Reykjavik, Iceland}, \bibinfo{pages}{423--431}.
\newblock
\showISBNx{979-8-3503-9511-2}
\href{https://doi.org/10.1109/RE59067.2024.00050}{doi:\nolinkurl{10.1109/RE59067.2024.00050}}


\bibitem[Belcak et~al\mbox{.}(2025)]%
        {belcakSmallLanguageModels2025}
\bibfield{author}{\bibinfo{person}{Peter Belcak}, \bibinfo{person}{Greg Heinrich}, \bibinfo{person}{Shizhe Diao}, \bibinfo{person}{Yonggan Fu}, \bibinfo{person}{Xin Dong}, \bibinfo{person}{Saurav Muralidharan}, \bibinfo{person}{Yingyan~Celine Lin}, {and} \bibinfo{person}{Pavlo Molchanov}.} \bibinfo{year}{2025}\natexlab{}.
\newblock \bibinfo{title}{Small {Language} {Models} are the {Future} of {Agentic} {AI}}.
\newblock
\href{https://doi.org/10.48550/arXiv.2506.02153}{doi:\nolinkurl{10.48550/arXiv.2506.02153}}


\bibitem[Bridgeford et~al\mbox{.}(2025)]%
        {bridgefordTenSimpleRules2025}
\bibfield{author}{\bibinfo{person}{Eric~W. Bridgeford}, \bibinfo{person}{Iain Campbell}, \bibinfo{person}{Zijao Chen}, \bibinfo{person}{Zhicheng Lin}, \bibinfo{person}{Harrison Ritz}, \bibinfo{person}{Joachim Vandekerckhove}, {and} \bibinfo{person}{Russell~A. Poldrack}.} \bibinfo{year}{2025}\natexlab{}.
\newblock \bibinfo{title}{Ten {Simple} {Rules} for {AI}-{Assisted} {Coding} in {Science}}.
\newblock
\href{https://doi.org/10.48550/arXiv.2510.22254}{doi:\nolinkurl{10.48550/arXiv.2510.22254}}


\bibitem[Burge and Brown(2008)]%
        {burgeSoftwareEngineeringUsing2008}
\bibfield{author}{\bibinfo{person}{Janet~E. Burge} {and} \bibinfo{person}{David~C. Brown}.} \bibinfo{year}{2008}\natexlab{}.
\newblock \showarticletitle{Software {Engineering} {Using} {RATionale}}.
\newblock \bibinfo{journal}{\emph{Journal of Systems and Software}} \bibinfo{volume}{81}, \bibinfo{number}{3} (\bibinfo{date}{March} \bibinfo{year}{2008}), \bibinfo{pages}{395--413}.
\newblock
\showISSN{0164-1212}
\href{https://doi.org/10.1016/j.jss.2007.05.004}{doi:\nolinkurl{10.1016/j.jss.2007.05.004}}


\bibitem[Carver(2012)]%
        {carverSoftwareEngineeringComputational2012}
\bibfield{author}{\bibinfo{person}{Jeffrey~C. Carver}.} \bibinfo{year}{2012}\natexlab{}.
\newblock \showarticletitle{Software {Engineering} for {Computational} {Science} and {Engineering}}.
\newblock \bibinfo{journal}{\emph{Computing in Science \& Engineering}} \bibinfo{volume}{14}, \bibinfo{number}{02} (\bibinfo{date}{March} \bibinfo{year}{2012}), \bibinfo{pages}{8--11}.
\newblock
\showISSN{1521-9615}
\href{https://doi.org/10.1109/MCSE.2012.31}{doi:\nolinkurl{10.1109/MCSE.2012.31}}


\bibitem[Conklin and Begeman(1988)]%
        {conklinGIBISHypertextTool1988}
\bibfield{author}{\bibinfo{person}{Jeff Conklin} {and} \bibinfo{person}{Michael~L. Begeman}.} \bibinfo{year}{1988}\natexlab{}.
\newblock \showarticletitle{{gIBIS}: {A} {Hypertext} {Tool} for {Exploratory} {Policy} {Discussion}}. In \bibinfo{booktitle}{\emph{Proceedings of the 1988 {ACM} conference on {Computer}-supported cooperative work}} \emph{(\bibinfo{series}{{CSCW} '88})}. \bibinfo{publisher}{Association for Computing Machinery}, \bibinfo{address}{New York, NY, USA}, \bibinfo{pages}{140--152}.
\newblock
\showISBNx{978-0-89791-282-2}
\href{https://doi.org/10.1145/62266.62278}{doi:\nolinkurl{10.1145/62266.62278}}


\bibitem[Dailey and Tambay(2024)]%
        {tambaySSECYearOneLearings2024}
\bibfield{author}{\bibinfo{person}{Dharma Dailey} {and} \bibinfo{person}{Anshul Tambay}.} \bibinfo{year}{2024}\natexlab{}.
\newblock \bibinfo{title}{Selected {Year} {One} {Learnings} of the {Scientific} {Software} {Engineering} {Center}}.
\newblock


\bibitem[Espinosa et~al\mbox{.}(2007)]%
        {espinosaFamiliarityComplexityTeam2007}
\bibfield{author}{\bibinfo{person}{J.~Alberto Espinosa}, \bibinfo{person}{Sandra~A. Slaughter}, \bibinfo{person}{Robert~E. Kraut}, {and} \bibinfo{person}{James~D. Herbsleb}.} \bibinfo{year}{2007}\natexlab{}.
\newblock \showarticletitle{Familiarity, {Complexity}, and {Team} {Performance} in {Geographically} {Distributed} {Software} {Development}}.
\newblock \bibinfo{journal}{\emph{Organization Science}} \bibinfo{volume}{18}, \bibinfo{number}{4} (\bibinfo{date}{Aug.} \bibinfo{year}{2007}), \bibinfo{pages}{613--630}.
\newblock
\showISSN{1047-7039}
\href{https://doi.org/10.1287/orsc.1070.0297}{doi:\nolinkurl{10.1287/orsc.1070.0297}}


\bibitem[Feng et~al\mbox{.}(2025)]%
        {fengLLMRouterOpenSourceLibrary2025}
\bibfield{author}{\bibinfo{person}{Tao Feng}, \bibinfo{person}{Haozhen Zhang}, \bibinfo{person}{Zijie Lei}, \bibinfo{person}{Haodong Yue}, \bibinfo{person}{Chongshan Lin}, \bibinfo{person}{Ge Liu}, {and} \bibinfo{person}{Jiaxuan You}.} \bibinfo{year}{2025}\natexlab{}.
\newblock \bibinfo{title}{{LLMRouter}: {An} {Open}-{Source} {Library} for {LLM} {Routing}}.
\newblock
\urldef\tempurl%
\url{https://github.com/ulab-uiuc/LLMRouter}
\showURL{%
\tempurl}


\bibitem[Gan and Sun(2025)]%
        {ganRAGMCPMitigatingPrompt2025}
\bibfield{author}{\bibinfo{person}{Tiantian Gan} {and} \bibinfo{person}{Qiyao Sun}.} \bibinfo{year}{2025}\natexlab{}.
\newblock \bibinfo{title}{{RAG}-{MCP}: {Mitigating} {Prompt} {Bloat} in {LLM} {Tool} {Selection} via {Retrieval}-{Augmented} {Generation}}.
\newblock
\href{https://doi.org/10.48550/arXiv.2505.03275}{doi:\nolinkurl{10.48550/arXiv.2505.03275}}


\bibitem[Gesing(2025)]%
        {gesingRSEs2035Surviving2025}
\bibfield{author}{\bibinfo{person}{Sandra Gesing}.} \bibinfo{year}{2025}\natexlab{}.
\newblock \showarticletitle{{RSEs} 2035: {Surviving} or {Thriving} in the {Age} of {AI}}. In \bibinfo{booktitle}{\emph{2025 {IEEE} {International} {Conference} on {eScience} ({eScience})}}. \bibinfo{pages}{381--382}.
\newblock
\showISSN{2325-3703}
\href{https://doi.org/10.1109/eScience65000.2025.00081}{doi:\nolinkurl{10.1109/eScience65000.2025.00081}}
\newblock
\shownote{ISSN: 2325-3703}.


\bibitem[Golia and Kalita(2024)]%
        {goliaActionItemDrivenSummarizationLong2024}
\bibfield{author}{\bibinfo{person}{Logan Golia} {and} \bibinfo{person}{Jugal Kalita}.} \bibinfo{year}{2024}\natexlab{}.
\newblock \showarticletitle{Action-{Item}-{Driven} {Summarization} of {Long} {Meeting} {Transcripts}}. In \bibinfo{booktitle}{\emph{Proceedings of the 2023 7th {International} {Conference} on {Natural} {Language} {Processing} and {Information} {Retrieval}}} \emph{(\bibinfo{series}{{NLPIR} '23})}. \bibinfo{publisher}{Association for Computing Machinery}, \bibinfo{address}{New York, NY, USA}, \bibinfo{pages}{91--98}.
\newblock
\showISBNx{979-8-4007-0922-7}
\href{https://doi.org/10.1145/3639233.3639253}{doi:\nolinkurl{10.1145/3639233.3639253}}


\bibitem[Gousios et~al\mbox{.}(2014)]%
        {gousiosExploratoryStudyPullbased2014}
\bibfield{author}{\bibinfo{person}{Georgios Gousios}, \bibinfo{person}{Martin Pinzger}, {and} \bibinfo{person}{Arie~van Deursen}.} \bibinfo{year}{2014}\natexlab{}.
\newblock \showarticletitle{An {Exploratory} {Study} of the {Pull}-{Based} {Software} {Development} {Model}}. In \bibinfo{booktitle}{\emph{Proceedings of the 36th {International} {Conference} on {Software} {Engineering}}} \emph{(\bibinfo{series}{{ICSE} 2014})}. \bibinfo{publisher}{Association for Computing Machinery}, \bibinfo{address}{New York, NY, USA}, \bibinfo{pages}{345--355}.
\newblock
\showISBNx{978-1-4503-2756-5}
\href{https://doi.org/10.1145/2568225.2568260}{doi:\nolinkurl{10.1145/2568225.2568260}}


\bibitem[Hannay et~al\mbox{.}(2009)]%
        {hannayHowScientistsDevelop2009}
\bibfield{author}{\bibinfo{person}{Jo~Erskine Hannay}, \bibinfo{person}{Carolyn MacLeod}, \bibinfo{person}{Janice Singer}, \bibinfo{person}{Hans~Petter Langtangen}, \bibinfo{person}{Dietmar Pfahl}, {and} \bibinfo{person}{Greg Wilson}.} \bibinfo{year}{2009}\natexlab{}.
\newblock \showarticletitle{How {Do} {Scientists} {Develop} and {Use} {Scientific} {Software}?}. In \bibinfo{booktitle}{\emph{2009 {ICSE} {Workshop} on {Software} {Engineering} for {Computational} {Science} and {Engineering}}}. \bibinfo{pages}{1--8}.
\newblock
\href{https://doi.org/10.1109/SECSE.2009.5069155}{doi:\nolinkurl{10.1109/SECSE.2009.5069155}}


\bibitem[Hata et~al\mbox{.}(2022)]%
        {hataGitHubDiscussionsExploratory2022}
\bibfield{author}{\bibinfo{person}{Hideaki Hata}, \bibinfo{person}{Nicole Novielli}, \bibinfo{person}{Sebastian Baltes}, \bibinfo{person}{Raula~Gaikovina Kula}, {and} \bibinfo{person}{Christoph Treude}.} \bibinfo{year}{2022}\natexlab{}.
\newblock \showarticletitle{{GitHub} {Discussions}: {An} {Exploratory} {Study} of {Early} {Adoption}}.
\newblock \bibinfo{journal}{\emph{Empirical Software Engineering}} \bibinfo{volume}{27}, \bibinfo{number}{1} (\bibinfo{date}{Jan.} \bibinfo{year}{2022}), \bibinfo{pages}{3}.
\newblock
\showISSN{1382-3256, 1573-7616}
\href{https://doi.org/10.1007/s10664-021-10058-6}{doi:\nolinkurl{10.1007/s10664-021-10058-6}}


\bibitem[Heaton and Carver(2015)]%
        {heatonClaimsUseSoftware2015}
\bibfield{author}{\bibinfo{person}{Dustin Heaton} {and} \bibinfo{person}{Jeffrey~C. Carver}.} \bibinfo{year}{2015}\natexlab{}.
\newblock \showarticletitle{Claims {About} the {Use} of {Software} {Engineering} {Practices} in {Science}: {A} {Systematic} {Literature} {Review}}.
\newblock \bibinfo{journal}{\emph{Information and Software Technology}}  \bibinfo{volume}{67} (\bibinfo{date}{Nov.} \bibinfo{year}{2015}), \bibinfo{pages}{207--219}.
\newblock
\showISSN{09505849}
\href{https://doi.org/10.1016/j.infsof.2015.07.011}{doi:\nolinkurl{10.1016/j.infsof.2015.07.011}}


\bibitem[Hocquet et~al\mbox{.}(2024)]%
        {hocquetnatcom2024}
\bibfield{author}{\bibinfo{person}{Alexandre Hocquet}, \bibinfo{person}{Frédéric Wieber}, \bibinfo{person}{Gabriele Gramelsberger}, \bibinfo{person}{Konrad Hinsen}, \bibinfo{person}{Markus Diesmann}, \bibinfo{person}{Fernando Pasquini~Santos}, \bibinfo{person}{Catharina Landström}, \bibinfo{person}{Benjamin Peters}, \bibinfo{person}{Dawid Kasprowicz}, \bibinfo{person}{Arianna Borrelli}, \bibinfo{person}{Phillip Roth}, \bibinfo{person}{Clarissa Ai~Ling Lee}, \bibinfo{person}{Alin Olteanu}, {and} \bibinfo{person}{Stefan Böschen}.} \bibinfo{year}{2024}\natexlab{}.
\newblock \showarticletitle{Software in {Science} {Is} {Ubiquitous} yet {Overlooked}}.
\newblock \bibinfo{journal}{\emph{Nature Computational Science}} (\bibinfo{date}{July} \bibinfo{year}{2024}), \bibinfo{pages}{1--4}.
\newblock
\showISSN{2662-8457}
\href{https://doi.org/10.1038/s43588-024-00651-2}{doi:\nolinkurl{10.1038/s43588-024-00651-2}}


\bibitem[Hofeditz et~al\mbox{.}(2024)]%
        {hofeditzEthicalChallengesHuman2024}
\bibfield{author}{\bibinfo{person}{Lennart Hofeditz}, \bibinfo{person}{Milad Mirbabaie}, {and} \bibinfo{person}{Mara Ortmann}.} \bibinfo{year}{2024}\natexlab{}.
\newblock \showarticletitle{Ethical {Challenges} for {Human}–{Agent} {Interaction} in {Virtual} {Collaboration} at {Work}}.
\newblock \bibinfo{journal}{\emph{International Journal of Human–Computer Interaction}} \bibinfo{volume}{40}, \bibinfo{number}{23} (\bibinfo{date}{Dec.} \bibinfo{year}{2024}), \bibinfo{pages}{8229--8245}.
\newblock
\showISSN{1044-7318, 1532-7590}
\href{https://doi.org/10.1080/10447318.2023.2279400}{doi:\nolinkurl{10.1080/10447318.2023.2279400}}


\bibitem[Horner and Atwood(2006)]%
        {hornerEffectiveDesignRationale2006}
\bibfield{author}{\bibinfo{person}{John Horner} {and} \bibinfo{person}{M.~E. Atwood}.} \bibinfo{year}{2006}\natexlab{}.
\newblock \showarticletitle{Effective {Design} {Rationale}: {Understanding} the {Barriers}}.
\newblock In \bibinfo{booktitle}{\emph{Rationale {Management} in {Software} {Engineering}}}, \bibfield{editor}{\bibinfo{person}{Allen~H. Dutoit}, \bibinfo{person}{Raymond McCall}, \bibinfo{person}{Ivan Mistrík}, {and} \bibinfo{person}{Barbara Paech}} (Eds.). \bibinfo{publisher}{Springer}, \bibinfo{address}{Berlin, Heidelberg}, \bibinfo{pages}{73--90}.
\newblock
\showISBNx{978-3-540-30998-7}
\href{https://doi.org/10.1007/978-3-540-30998-7_3}{doi:\nolinkurl{10.1007/978-3-540-30998-7_3}}


\bibitem[Hosseini et~al\mbox{.}(2024)]%
        {hosseiniOpenScienceGenerative2024}
\bibfield{author}{\bibinfo{person}{Mohammad Hosseini}, \bibinfo{person}{Serge P J~M Horbach}, \bibinfo{person}{Kristi~L Holmes}, {and} \bibinfo{person}{Tony Ross-Hellauer}.} \bibinfo{year}{2024}\natexlab{}.
\newblock \bibinfo{title}{Open {Science} at the {Generative} {AI} {Turn}: {An} {Exploratory} {Analysis} of {Challenges} and {Opportunities}}.
\newblock
\href{https://doi.org/10.1162/qss_a_00337}{doi:\nolinkurl{10.1162/qss_a_00337}}


\bibitem[Hunter-Zinck et~al\mbox{.}(2021)]%
        {hunter-zinckTenSimpleRules2021}
\bibfield{author}{\bibinfo{person}{Haley Hunter-Zinck}, \bibinfo{person}{Alexandre~Fioravante De~Siqueira}, \bibinfo{person}{Váleri~N. Vásquez}, \bibinfo{person}{Richard Barnes}, {and} \bibinfo{person}{Ciera~C. Martinez}.} \bibinfo{year}{2021}\natexlab{}.
\newblock \showarticletitle{Ten {Simple} {Rules} on {Writing} {Clean} and {Reliable} {Open}-{Source} {Scientific} {Software}}.
\newblock \bibinfo{journal}{\emph{PLOS Computational Biology}} \bibinfo{volume}{17}, \bibinfo{number}{11} (\bibinfo{date}{Nov.} \bibinfo{year}{2021}), \bibinfo{pages}{e1009481}.
\newblock
\showISSN{1553-7358}
\href{https://doi.org/10.1371/journal.pcbi.1009481}{doi:\nolinkurl{10.1371/journal.pcbi.1009481}}


\bibitem[Jimenez et~al\mbox{.}(2023)]%
        {jimenezSWEbenchCanLanguage2023}
\bibfield{author}{\bibinfo{person}{Carlos~E. Jimenez}, \bibinfo{person}{John Yang}, \bibinfo{person}{Alexander Wettig}, \bibinfo{person}{Shunyu Yao}, \bibinfo{person}{Kexin Pei}, \bibinfo{person}{Ofir Press}, {and} \bibinfo{person}{Karthik~R. Narasimhan}.} \bibinfo{year}{2023}\natexlab{}.
\newblock \showarticletitle{{SWE}-{Bench}: {Can} {Language} {Models} {Resolve} {Real}-{World} {GitHub} {Issues}?}
\newblock
\urldef\tempurl%
\url{https://openreview.net/forum?id=VTF8yNQM66}
\showURL{%
\tempurl}


\bibitem[Johanson and Hasselbring(2018)]%
        {johansonSoftwareEngineeringComputational2018}
\bibfield{author}{\bibinfo{person}{Arne Johanson} {and} \bibinfo{person}{Wilhelm Hasselbring}.} \bibinfo{year}{2018}\natexlab{}.
\newblock \showarticletitle{Software {Engineering} for {Computational} {Science}: {Past}, {Present}, {Future}}.
\newblock \bibinfo{journal}{\emph{Computing in Science \& Engineering}} \bibinfo{volume}{20}, \bibinfo{number}{2} (\bibinfo{date}{March} \bibinfo{year}{2018}), \bibinfo{pages}{90--109}.
\newblock
\showISSN{1521-9615, 1558-366X}
\href{https://doi.org/10.1109/MCSE.2018.021651343}{doi:\nolinkurl{10.1109/MCSE.2018.021651343}}


\bibitem[Kabir et~al\mbox{.}(2024)]%
        {kabirStackOverflowObsolete2024}
\bibfield{author}{\bibinfo{person}{Samia Kabir}, \bibinfo{person}{David~N. Udo-Imeh}, \bibinfo{person}{Bonan Kou}, {and} \bibinfo{person}{Tianyi Zhang}.} \bibinfo{year}{2024}\natexlab{}.
\newblock \showarticletitle{Is {Stack} {Overflow} {Obsolete}? {An} {Empirical} {Study} of the {Characteristics} of {ChatGPT} {Answers} to {Stack} {Overflow} {Questions}}. In \bibinfo{booktitle}{\emph{Proceedings of the 2024 {CHI} {Conference} on {Human} {Factors} in {Computing} {Systems}}} \emph{(\bibinfo{series}{{CHI} '24})}. \bibinfo{publisher}{Association for Computing Machinery}, \bibinfo{address}{New York, NY, USA}, \bibinfo{pages}{1--17}.
\newblock
\showISBNx{979-8-4007-0330-0}
\href{https://doi.org/10.1145/3613904.3642596}{doi:\nolinkurl{10.1145/3613904.3642596}}


\bibitem[Laban et~al\mbox{.}(2025)]%
        {labanLLMsGetLost2025}
\bibfield{author}{\bibinfo{person}{Philippe Laban}, \bibinfo{person}{Hiroaki Hayashi}, \bibinfo{person}{Yingbo Zhou}, {and} \bibinfo{person}{Jennifer Neville}.} \bibinfo{year}{2025}\natexlab{}.
\newblock \bibinfo{title}{{LLMs} {Get} {Lost} {In} {Multi}-{Turn} {Conversation}}.
\newblock
\href{https://doi.org/10.48550/arXiv.2505.06120}{doi:\nolinkurl{10.48550/arXiv.2505.06120}}


\bibitem[Labs({[n.\,d.]})]%
        {PaperclipAppPeople}
\bibfield{author}{\bibinfo{person}{Paperclip Labs}.} \bibinfo{year}{[n.\,d.]}\natexlab{}.
\newblock \bibinfo{title}{Paperclip – {The} {App} {People} {Use} to {Manage} {AI} {Agents} for {Work}}.
\newblock
\urldef\tempurl%
\url{https://paperclip.ing/}
\showURL{%
\tempurl}


\bibitem[Lee(2007)]%
        {leeBoundaryNegotiatingArtifacts2007}
\bibfield{author}{\bibinfo{person}{Charlotte~P. Lee}.} \bibinfo{year}{2007}\natexlab{}.
\newblock \showarticletitle{Boundary {Negotiating} {Artifacts}: {Unbinding} the {Routine} of {Boundary} {Objects} and {Embracing} {Chaos} in {Collaborative} {Work}}.
\newblock \bibinfo{journal}{\emph{Computer Supported Cooperative Work (CSCW)}} \bibinfo{volume}{16}, \bibinfo{number}{3} (\bibinfo{date}{June} \bibinfo{year}{2007}), \bibinfo{pages}{307--339}.
\newblock
\showISSN{0925-9724}
\href{https://doi.org/10.1007/s10606-007-9044-5}{doi:\nolinkurl{10.1007/s10606-007-9044-5}}


\bibitem[Li et~al\mbox{.}(2025)]%
        {liRiseAITeammates2025}
\bibfield{author}{\bibinfo{person}{Hao Li}, \bibinfo{person}{Haoxiang Zhang}, {and} \bibinfo{person}{Ahmed~E. Hassan}.} \bibinfo{year}{2025}\natexlab{}.
\newblock \bibinfo{title}{The {Rise} of {AI} {Teammates} in {Software} {Engineering} ({SE}) 3.0: {How} {Autonomous} {Coding} {Agents} {Are} {Reshaping} {Software} {Engineering}}.
\newblock
\href{https://doi.org/10.48550/arXiv.2507.15003}{doi:\nolinkurl{10.48550/arXiv.2507.15003}}


\bibitem[Li et~al\mbox{.}(2026)]%
        {liLLMRouterBenchMassiveBenchmark2026}
\bibfield{author}{\bibinfo{person}{Hao Li}, \bibinfo{person}{Yiqun Zhang}, \bibinfo{person}{Zhaoyan Guo}, \bibinfo{person}{Chenxu Wang}, \bibinfo{person}{Shengji Tang}, \bibinfo{person}{Qiaosheng Zhang}, \bibinfo{person}{Yang Chen}, \bibinfo{person}{Biqing Qi}, \bibinfo{person}{Peng Ye}, \bibinfo{person}{Lei Bai}, \bibinfo{person}{Zhen Wang}, {and} \bibinfo{person}{Shuyue Hu}.} \bibinfo{year}{2026}\natexlab{}.
\newblock \showarticletitle{{LLMRouterBench}: {A} {Massive} {Benchmark} and {Unified} {Framework} for {LLM} {Routing}}.
\newblock  (\bibinfo{year}{2026}).
\newblock
\href{https://doi.org/10.48550/ARXIV.2601.07206}{doi:\nolinkurl{10.48550/ARXIV.2601.07206}}


\bibitem[McInnes et~al\mbox{.}(2025)]%
        {mcinnesReport2025Workshop2025}
\bibfield{author}{\bibinfo{person}{Lois~Curfman McInnes}, \bibinfo{person}{Dorian Arnold}, \bibinfo{person}{Prasanna Balaprakash}, \bibinfo{person}{Mike Bernhardt}, \bibinfo{person}{Beth Cerny}, \bibinfo{person}{Anshu Dubey}, \bibinfo{person}{Roscoe Giles}, \bibinfo{person}{Denice~Ward Hood}, \bibinfo{person}{Mary~Ann Leung}, \bibinfo{person}{Vanessa Lopez-Marrero}, \bibinfo{person}{Paul Messina}, \bibinfo{person}{Olivia~B. Newton}, \bibinfo{person}{Chris Oehmen}, \bibinfo{person}{Stefan~M. Wild}, \bibinfo{person}{Jim Willenbring}, \bibinfo{person}{Lou Woodley}, \bibinfo{person}{Tony Baylis}, \bibinfo{person}{David~E. Bernholdt}, \bibinfo{person}{Chris Camano}, \bibinfo{person}{Johannah Cohoon}, \bibinfo{person}{Charles Ferenbaugh}, \bibinfo{person}{Stephen~M. Fiore}, \bibinfo{person}{Sandra Gesing}, \bibinfo{person}{Diego Gomez-Zara}, \bibinfo{person}{James Howison}, \bibinfo{person}{Tanzima Islam}, \bibinfo{person}{David Kepczynski}, \bibinfo{person}{Charles Lively}, \bibinfo{person}{Harshitha Menon},
  \bibinfo{person}{Bronson Messer}, \bibinfo{person}{Marieme Ngom}, \bibinfo{person}{Umesh Paliath}, \bibinfo{person}{Michael~E. Papka}, \bibinfo{person}{Irene Qualters}, \bibinfo{person}{Elaine~M. Raybourn}, \bibinfo{person}{Katherine Riley}, \bibinfo{person}{Paulina Rodriguez}, \bibinfo{person}{Damian Rouson}, \bibinfo{person}{Michelle Schwalbe}, \bibinfo{person}{Sudip~K. Seal}, \bibinfo{person}{Ozge Surer}, \bibinfo{person}{Valerie Taylor}, {and} \bibinfo{person}{Lingfei Wu}.} \bibinfo{year}{2025}\natexlab{}.
\newblock \bibinfo{title}{Report of the 2025 {Workshop} on {Next}-{Generation} {Ecosystems} for {Scientific} {Computing}: {Harnessing} {Community}, {Software}, and {AI} for {Cross}-{Disciplinary} {Team} {Science}}.
\newblock
\href{https://doi.org/10.48550/arXiv.2510.03413}{doi:\nolinkurl{10.48550/arXiv.2510.03413}}


\bibitem[Nieuwpoort and Katz(2023)]%
        {nieuwpoortDefiningRolesResearch2023}
\bibfield{author}{\bibinfo{person}{Rob~van Nieuwpoort} {and} \bibinfo{person}{Daniel~S. Katz}.} \bibinfo{year}{2023}\natexlab{}.
\newblock \showarticletitle{Defining the {Roles} of {Research} {Software}}.
\newblock \bibinfo{journal}{\emph{Upstream}} (\bibinfo{date}{March} \bibinfo{year}{2023}).
\newblock
\href{https://doi.org/10.54900/9akm9y5-5ject5y}{doi:\nolinkurl{10.54900/9akm9y5-5ject5y}}


\bibitem[{Notion Labs, Inc.}(2026)]%
        {Notion}
\bibfield{author}{\bibinfo{person}{{Notion Labs, Inc.}}} \bibinfo{year}{2026}\natexlab{}.
\newblock \bibinfo{title}{Notion}.
\newblock
\urldef\tempurl%
\url{https://www.notion.com/}
\showURL{%
\tempurl}


\bibitem[O'Brien(2025)]%
        {obrienHowScientistsUse2025}
\bibfield{author}{\bibinfo{person}{Gabrielle O'Brien}.} \bibinfo{year}{2025}\natexlab{}.
\newblock \showarticletitle{How {Scientists} {Use} {Large} {Language} {Models} to {Program}}. In \bibinfo{booktitle}{\emph{Proceedings of the 2025 {CHI} {Conference} on {Human} {Factors} in {Computing} {Systems}}} \emph{(\bibinfo{series}{{CHI} '25})}. \bibinfo{publisher}{Association for Computing Machinery}, \bibinfo{address}{New York, NY, USA}, \bibinfo{pages}{1--16}.
\newblock
\showISBNx{979-8-4007-1394-1}
\href{https://doi.org/10.1145/3706598.3713668}{doi:\nolinkurl{10.1145/3706598.3713668}}


\bibitem[O'Brien et~al\mbox{.}(2026)]%
        {obrienSurveyGenerativeAI2026}
\bibfield{author}{\bibinfo{person}{Gabrielle O'Brien}, \bibinfo{person}{Alexis Parker}, \bibinfo{person}{Nasir Eisty}, {and} \bibinfo{person}{Jeffrey Carver}.} \bibinfo{year}{2026}\natexlab{}.
\newblock \bibinfo{title}{A {Survey} of {Generative} {AI} {Adoption} and {Perceived} {Productivity} {Among} {Scientists} {Who} {Program}}.
\newblock
\href{https://doi.org/10.48550/arXiv.2512.19644}{doi:\nolinkurl{10.48550/arXiv.2512.19644}}


\bibitem[O’Brien(2025)]%
        {obrienThreatsScientificSoftware2025}
\bibfield{author}{\bibinfo{person}{Gabrielle O’Brien}.} \bibinfo{year}{2025}\natexlab{}.
\newblock \showarticletitle{Threats to {Scientific} {Software} {From} {Over}-{Reliance} on {AI} {Code} {Assistants}}.
\newblock \bibinfo{journal}{\emph{Nature Computational Science}} \bibinfo{volume}{5}, \bibinfo{number}{9} (\bibinfo{date}{Sept.} \bibinfo{year}{2025}), \bibinfo{pages}{701--703}.
\newblock
\showISSN{2662-8457}
\href{https://doi.org/10.1038/s43588-025-00845-2}{doi:\nolinkurl{10.1038/s43588-025-00845-2}}


\bibitem[Pham et~al\mbox{.}(2017)]%
        {phamOnboardingInexperiencedDevelopers2017}
\bibfield{author}{\bibinfo{person}{Raphael Pham}, \bibinfo{person}{Stephan Kiesling}, \bibinfo{person}{Leif Singer}, {and} \bibinfo{person}{Kurt Schneider}.} \bibinfo{year}{2017}\natexlab{}.
\newblock \showarticletitle{Onboarding {Inexperienced} {Developers}: {Struggles} and {Perceptions} {Regarding} {Automated} {Testing}}.
\newblock \bibinfo{journal}{\emph{Software Quality Journal}} \bibinfo{volume}{25}, \bibinfo{number}{4} (\bibinfo{date}{Dec.} \bibinfo{year}{2017}), \bibinfo{pages}{1239--1268}.
\newblock
\showISSN{0963-9314}
\href{https://doi.org/10.1007/s11219-016-9333-7}{doi:\nolinkurl{10.1007/s11219-016-9333-7}}


\bibitem[Research(2026)]%
        {NousResearchHermesagent2026}
\bibfield{author}{\bibinfo{person}{Nous Research}.} \bibinfo{year}{2026}\natexlab{}.
\newblock \bibinfo{title}{Hermes {Agent}}.
\newblock
\urldef\tempurl%
\url{https://github.com/NousResearch/hermes-agent}
\showURL{%
\tempurl}


\bibitem[Salas and Converse(1993)]%
        {salasSharedMentalModels1993}
\bibfield{author}{\bibinfo{person}{Janis A. Cannon-Bowers~Eduardo Salas} {and} \bibinfo{person}{Sharolyn Converse}.} \bibinfo{year}{1993}\natexlab{}.
\newblock \showarticletitle{Shared {Mental} {Models} in {Expert} {Team} {Decision} {Making}*}.
\newblock In \bibinfo{booktitle}{\emph{Individual and {Group} {Decision} {Making}}}. \bibinfo{publisher}{Psychology Press}.
\newblock
\newblock
\shownote{Num Pages: 26}.


\bibitem[Sanders and Kelly(2008)]%
        {sandersDealingRiskScientific2008}
\bibfield{author}{\bibinfo{person}{Rebecca Sanders} {and} \bibinfo{person}{Diane Kelly}.} \bibinfo{year}{2008}\natexlab{}.
\newblock \showarticletitle{Dealing {With} {Risk} in {Scientific} {Software} {Development}}.
\newblock \bibinfo{journal}{\emph{IEEE Software}} \bibinfo{volume}{25}, \bibinfo{number}{4} (\bibinfo{date}{July} \bibinfo{year}{2008}), \bibinfo{pages}{21--28}.
\newblock
\showISSN{1937-4194}
\href{https://doi.org/10.1109/MS.2008.84}{doi:\nolinkurl{10.1109/MS.2008.84}}


\bibitem[Schick et~al\mbox{.}(2023)]%
        {schickToolformerLanguageModels2023}
\bibfield{author}{\bibinfo{person}{Timo Schick}, \bibinfo{person}{Jane Dwivedi-Yu}, \bibinfo{person}{Roberto Dessi}, \bibinfo{person}{Roberta Raileanu}, \bibinfo{person}{Maria Lomeli}, \bibinfo{person}{Eric Hambro}, \bibinfo{person}{Luke Zettlemoyer}, \bibinfo{person}{Nicola Cancedda}, {and} \bibinfo{person}{Thomas Scialom}.} \bibinfo{year}{2023}\natexlab{}.
\newblock \showarticletitle{Toolformer: {Language} {Models} {Can} {Teach} {Themselves} {To} {Use} {Tools}}.
\newblock
\urldef\tempurl%
\url{https://openreview.net/forum?id=Yacmpz84TH}
\showURL{%
\tempurl}


\bibitem[Shen and Tamkin(2026)]%
        {shenHowAIImpacts2026}
\bibfield{author}{\bibinfo{person}{Judy~Hanwen Shen} {and} \bibinfo{person}{Alex Tamkin}.} \bibinfo{year}{2026}\natexlab{}.
\newblock \bibinfo{title}{How {AI} {Impacts} {Skill} {Formation}}.
\newblock
\href{https://doi.org/10.48550/arXiv.2601.20245}{doi:\nolinkurl{10.48550/arXiv.2601.20245}}


\bibitem[Star and Griesemer(1989)]%
        {starInstitutionalEcologyTranslations1989}
\bibfield{author}{\bibinfo{person}{Susan~Leigh Star} {and} \bibinfo{person}{James~R. Griesemer}.} \bibinfo{year}{1989}\natexlab{}.
\newblock \showarticletitle{Institutional {Ecology}, ‘{Translations}’ and {Boundary} {Objects}: {Amateurs} and {Professionals} in {Berkeley}’s {Museum} of {Vertebrate} {Zoology}, 1907–39}.
\newblock \bibinfo{journal}{\emph{Social Studies of Science}} \bibinfo{volume}{19}, \bibinfo{number}{3} (\bibinfo{date}{Aug.} \bibinfo{year}{1989}), \bibinfo{pages}{387--420}.
\newblock
\showISSN{0306-3127}
\href{https://doi.org/10.1177/030631289019003001}{doi:\nolinkurl{10.1177/030631289019003001}}


\bibitem[Steinberger({[n.\,d.]})]%
        {OpenClawPersonalAI}
\bibfield{author}{\bibinfo{person}{Peter Steinberger}.} \bibinfo{year}{[n.\,d.]}\natexlab{}.
\newblock \bibinfo{title}{{OpenClaw} — {Personal} {AI} {Assistant}}.
\newblock
\urldef\tempurl%
\url{https://openclaw.ai/}
\showURL{%
\tempurl}


\bibitem[Storey(2026)]%
        {storeyTechnicalDebtCognitive2026}
\bibfield{author}{\bibinfo{person}{Margaret-Anne Storey}.} \bibinfo{year}{2026}\natexlab{}.
\newblock \bibinfo{title}{From {Technical} {Debt} to {Cognitive} and {Intent} {Debt}: {Rethinking} {Software} {Health} in the {Age} of {AI}}.
\newblock
\href{https://doi.org/10.48550/arXiv.2603.22106}{doi:\nolinkurl{10.48550/arXiv.2603.22106}}


\bibitem[Storey and Zagalsky(2016)]%
        {storeyDisruptingDeveloperProductivity2016}
\bibfield{author}{\bibinfo{person}{Margaret-Anne Storey} {and} \bibinfo{person}{Alexey Zagalsky}.} \bibinfo{year}{2016}\natexlab{}.
\newblock \showarticletitle{Disrupting {Developer} {Productivity} {One} {Bot} at a {Time}}. In \bibinfo{booktitle}{\emph{Proceedings of the 2016 24th {ACM} {SIGSOFT} {International} {Symposium} on {Foundations} of {Software} {Engineering}}} \emph{(\bibinfo{series}{{FSE} 2016})}. \bibinfo{publisher}{Association for Computing Machinery}, \bibinfo{address}{New York, NY, USA}, \bibinfo{pages}{928--931}.
\newblock
\showISBNx{978-1-4503-4218-6}
\href{https://doi.org/10.1145/2950290.2983989}{doi:\nolinkurl{10.1145/2950290.2983989}}


\bibitem[Stray and Moe(2020)]%
        {strayUnderstandingCoordinationGlobal2020}
\bibfield{author}{\bibinfo{person}{Viktoria Stray} {and} \bibinfo{person}{Nils~Brede Moe}.} \bibinfo{year}{2020}\natexlab{}.
\newblock \showarticletitle{Understanding {Coordination} in {Global} {Software} {Engineering}: {A} {Mixed}-{Methods} {Study} on the {Use} of {Meetings} and {Slack}}.
\newblock \bibinfo{journal}{\emph{Journal of Systems and Software}}  \bibinfo{volume}{170} (\bibinfo{date}{Dec.} \bibinfo{year}{2020}), \bibinfo{pages}{110717}.
\newblock
\showISSN{01641212}
\href{https://doi.org/10.1016/j.jss.2020.110717}{doi:\nolinkurl{10.1016/j.jss.2020.110717}}


\bibitem[Tsay et~al\mbox{.}(2014)]%
        {tsayInfluenceSocialTechnical2014}
\bibfield{author}{\bibinfo{person}{Jason Tsay}, \bibinfo{person}{Laura Dabbish}, {and} \bibinfo{person}{James Herbsleb}.} \bibinfo{year}{2014}\natexlab{}.
\newblock \showarticletitle{Influence of {Social} and {Technical} {Factors} for {Evaluating} {Contribution} in {GitHub}}. In \bibinfo{booktitle}{\emph{Proceedings of the 36th {International} {Conference} on {Software} {Engineering}}} \emph{(\bibinfo{series}{{ICSE} 2014})}. \bibinfo{publisher}{Association for Computing Machinery}, \bibinfo{address}{New York, NY, USA}, \bibinfo{pages}{356--366}.
\newblock
\showISBNx{978-1-4503-2756-5}
\href{https://doi.org/10.1145/2568225.2568315}{doi:\nolinkurl{10.1145/2568225.2568315}}


\bibitem[Wang et~al\mbox{.}(2024)]%
        {wangSurveyLargeLanguage2024}
\bibfield{author}{\bibinfo{person}{Lei Wang}, \bibinfo{person}{Chen Ma}, \bibinfo{person}{Xueyang Feng}, \bibinfo{person}{Zeyu Zhang}, \bibinfo{person}{Hao Yang}, \bibinfo{person}{Jingsen Zhang}, \bibinfo{person}{Zhiyuan Chen}, \bibinfo{person}{Jiakai Tang}, \bibinfo{person}{Xu Chen}, \bibinfo{person}{Yankai Lin}, \bibinfo{person}{Wayne~Xin Zhao}, \bibinfo{person}{Zhewei Wei}, {and} \bibinfo{person}{Ji-Rong Wen}.} \bibinfo{year}{2024}\natexlab{}.
\newblock \showarticletitle{A {Survey} on {Large} {Language} {Model} {Based} {Autonomous} {Agents}}.
\newblock \bibinfo{journal}{\emph{Frontiers of Computer Science}} \bibinfo{volume}{18}, \bibinfo{number}{6} (\bibinfo{date}{Dec.} \bibinfo{year}{2024}), \bibinfo{pages}{186345}.
\newblock
\showISSN{2095-2228, 2095-2236}
\href{https://doi.org/10.1007/s11704-024-40231-1}{doi:\nolinkurl{10.1007/s11704-024-40231-1}}


\bibitem[Wu et~al\mbox{.}(2025)]%
        {wuHumangenerativeAICollaboration2025}
\bibfield{author}{\bibinfo{person}{Suqing Wu}, \bibinfo{person}{Yukun Liu}, \bibinfo{person}{Mengqi Ruan}, \bibinfo{person}{Siyu Chen}, {and} \bibinfo{person}{Xiao-Yun Xie}.} \bibinfo{year}{2025}\natexlab{}.
\newblock \showarticletitle{Human-{Generative} {AI} {Collaboration} {Enhances} {Task} {Performance} but {Undermines} {Human}’s {Intrinsic} {Motivation}}.
\newblock \bibinfo{journal}{\emph{Scientific Reports}} \bibinfo{volume}{15}, \bibinfo{number}{1} (\bibinfo{date}{April} \bibinfo{year}{2025}), \bibinfo{pages}{15105}.
\newblock
\showISSN{2045-2322}
\href{https://doi.org/10.1038/s41598-025-98385-2}{doi:\nolinkurl{10.1038/s41598-025-98385-2}}


\bibitem[Yang et~al\mbox{.}(2024)]%
        {yangSWEagentAgentComputerInterfaces2024}
\bibfield{author}{\bibinfo{person}{John Yang}, \bibinfo{person}{Carlos Jimenez}, \bibinfo{person}{Alexander Wettig}, \bibinfo{person}{Kilian Lieret}, \bibinfo{person}{Shunyu Yao}, \bibinfo{person}{Karthik Narasimhan}, {and} \bibinfo{person}{Ofir Press}.} \bibinfo{year}{2024}\natexlab{}.
\newblock \showarticletitle{{SWE}-agent: {Agent}-{Computer} {Interfaces} {Enable} {Automated} {Software} {Engineering}}.
\newblock \bibinfo{journal}{\emph{Advances in Neural Information Processing Systems}}  \bibinfo{volume}{37} (\bibinfo{date}{Dec.} \bibinfo{year}{2024}), \bibinfo{pages}{50528--50652}.
\newblock
\href{https://doi.org/10.52202/079017-1601}{doi:\nolinkurl{10.52202/079017-1601}}


\end{thebibliography}
